\documentstyle[12pt]{article} 
\begin{document}
\def \lsim {~\mbox{${}^< \hspace*{-7pt} _\sim$}~}
\def \gsim {~\mbox{${}^> \hspace*{-7pt} _\sim$}~}
\def \leql {~\ ^< \hspace*{-7pt} _=~\ }
\def \geql {~\ ^> \hspace*{-7pt} _=~\ }
\def\preRate{\Bigl[ N \frac{\rho_\chi}{m_\chi} \Bigr]}
\def\Rate0{\Bigl[ N \frac{\rho_\chi}{m_\chi} \sigma(0)
                  \langle v \rangle \Bigr]}
\def\dsdq2{\frac{d\sigma}{dq^2}}
\def\vd{\bar{v}}
\def\sqpi{\sqrt{\pi}}
\def\erf{\mbox{erf}}
\def\and{\mbox{~~~and~~~}}
\newcommand{\be}[1]{\begin{equation} \label{(#1)}}
\newcommand{\ee}{\end{equation}}
\newcommand{\ba}[1]{\begin{eqnarray} \label{(#1)}}
\newcommand{\ea}{\end{eqnarray}}
\newcommand{\nn}{\nonumber}
\newcommand{\rf}[1]{(\ref{(#1)})}
\newcommand{\bq}{\begin{equation}}
\newcommand{\eq}{\end{equation}}
\newcommand{\beq}  {\begin{eqnarray}}
\newcommand{\eeq}  {\end{eqnarray}}
\newcommand{\mt}   {{\ifmmode m_{t}         \else $m_{t}$          \fi}}
\newcommand{\tb}   {{\ifmmode \tan\beta     \else $\tan\beta$      \fi}}
\newcommand{\mz}   {{\ifmmode M_{Z}         \else $M_{Z}$          \fi}}
\newcommand{\bsg}  {{\ifmmode \b\rightarrow s\gamma
                        \else $b\rightarrow s\gamma$ \fi}}
\newcommand{\Bbsg}  {{\ifmmode BR(\b\rightarrow s\gamma)
                        \else $BR(b\rightarrow s\gamma)$ \fi}}
\newcommand{\smas}[2]{m^#2_{\tilde #1}}
\renewcommand{\floatpagefraction}{0.005}
\newcommand{\Zto}{\mbox{$\mathrm Z \to$}}
\setlength{\unitlength}{1in}

\begin{center}
{\Large\bf Is SUSY accessible by direct dark matter detection? }

\vspace{0.3cm}

        V.A.Bednyakov\footnotemark[1],
        S.G.Kovalenko\footnotemark[2] \\
        {\it
        Laboratory of Nuclear Problems,
        Joint Institute for Nuclear Research, \\
        Moscow region, 141980 Dubna, Russia
        } \\
        and \\
        H.V.Klapdor-Kleingrothaus\footnotemark[3],
        Y.Ramachers\footnotemark[4] \\
        {\it
        Max-Planck-Institut f\"{u}r Kernphysik,
        Postfach 103980, D-69029, Heidelberg, Germany
       }

\footnotetext[1]{E-mail: bedny@nusun.jinr.ru}
\footnotetext[2]{E-mail: kovalen@nusun.jinr.ru}
\footnotetext[3]{E-mail: klapdor@enull.mpi-hd.mpg.de}
\footnotetext[4]{E-mail: yorck@mickey.mpi-hd.mpg.de}

\vspace*{1cm}

{\bf Abstract}

\end{center}

        We performed a combined analysis of the parameter space
        of the Minimal Supersymmetric Standard Model (MSSM) taking
	into account cosmological and accelerator constraints
	including those from the radiative \bsg decay measured 
	 by the CLEO collaboration.

        Special attention is paid to the event rate, $R$,
        of direct dark matter neutralino detection.
        We have found domains of the parameter space with 
        $R\simeq 5-10$~events/kg/day.
        This would be  within the reach of current dark matter experiments.
        The \bsg data do not essentially reduce these large event rate
        domains of the MSSM parameter space.

\newpage

\section{Introduction}
        Rare \bsg decay  observed by the CLEO collaboration
\cite{cleo94} 
        with the branching ratio
        $ Br(b\to s\gamma) = (2.32 \pm 0.67)\cdot 10^{-4}$,
        has been recognized as a stringent 
	restriction for physics beyond the standard model (SM).

   	In the Minimal Supersymmetric Standard Model (MSSM)
\cite{rev} 
   	this decay proceeds through 1-loop diagrams involving  
	W-boson,  charged Higgs boson, chargino, neutralino and gluino
\cite{BerBorMasRi,BarbGiud}.
   	Since the SM prediction is consistent with the measured
	branching ratio \Bbsg the MSSM contributions are stringently 
	restricted. 
        A dramatic reduction of the allowed MSSM parameter
        space due to the $b\to s\gamma$ constraint was reported
        by many authors
\cite{BorDrNoj,Drees_bsg}.
	The impact of this constraint on prospects for direct
	detection of the dark matter (DM) neutralino ($\chi$)
        via elastic scattering off various nuclei has been also analyzed.
        In 
\cite{BorDrNoj,Drees_bsg}
        it was found that within  popular supergravity models
        the detection rate becomes too small for observation if 
	the CLEO constraint is incorporated in the analysis.
        In this case the upper bound on \Bbsg implies a stringent
        lower bound  on the mass of the pseudoscalar Higgs boson
        ($m_A$) of the MSSM if sparticles are heavy.
        It leads to strong suppression
        of the elastic neutralino-nucleus scattering cross section.

        Scenarios with lighter sparticles and pseudoscalar Higgs
        have been studied in the literature as well
\cite{Drees_bsg}.
        Also here, first results for direct detection of 
	neutralinos were pessimistic.
        Again, only small counting rates had been found in the domain
        of the MSSM parameter space satisfying the \Bbsg constraint.

        A more comprehensive exploration of the constrained MSSM 
	parameter space
\cite{roskane,BBEFMS,JuKaGr}
        discovered, however, that the \bsg constraint makes actually
        only a moderate effect on the expected event rate.
        It was realized that relatively light neutral Higgs bosons,
        leading to typically large event rates, are compatible with 
	this constraint in contrast with the results of previous analyses.
        A key observation is that the charged Higgs boson contribution
        to \Bbsg  can be efficiently compensated by the chargino
        contribution in a large domain of the MSSM parameter space.
        This essentially relaxes the lower bound both for
        charged and for neutral Higgses imposed by the CLEO result.

        Recently an exciting result was obtained in 
\cite{GonBer}.
        A sophisticated scan of the MSSM parameter space constrained
        by the known experimental bounds including \Bbsg picked up 
	points with an unexpectedly large detection rate of DM 
	neutralinos with a mass around 1~TeV.
        For a germanium ($^{76}$Ge) target an integrated detection rate 
	was found at a level of 10~events/kg/day and even up to 
	100~events/kg/day for sodium iodide (NaI).

        If these results are correct they have important consequences
        for direct DM detection.
	Such large allowed event rates would mean that the current
        DM experiments have already entered an unexplored part of 
	the MSSM parameter space.

        Certainly, before such a conclusion can be made one has to be sure
        that the above cited results, obtained in a specially arranged scan,
        are not an artifact having no relation to physics.
        One may suspect, for instance, a specific instability of the
        numerical code used in the analysis.
        Note, that the standard scan without special sampling of the
        model parameters did not produce even a single point with an 
	event rate larger than 1~event/kg/day.
        Therefore, an independent search for large event rate
        points within the MSSM parameter space is apparently demanded.

\smallskip

        In the present paper we carry out a systematic scan of the 
	MSSM parameter space constrained by the known accelerator
        data and by the requirement that the DM neutralinos do not 
	overclose the universe.
        We adopt the unification scenario
\cite{BKK}
        with a non-universal scalar mass when the soft Higgs mass 
	parameters are not equal to the common sfermion soft mass 
	parameter at the unification scale.
        In this case the Higgs and sfermion masses are not strongly 
	correlated parameters.
        As discussed in 
\cite{BKK}, this minimal relaxation of the complete unification
        conditions allows one to avoid one of the most stringent
        theoretical limitations on the allowed values of the neutralino
        detection event rate.
        Other unification conditions do not make such an effect and 
	tolerate large event rate values.
        Therefore, we keep these unification conditions to reduce
        a number of free parameters.
        The latter is crucial for a fine scanning of the MSSM parameter
        space which we are going to carry out.

        The aim of the scan is to "detect" those domains in the parameter
        space where the event rate $R$\ of the direct DM detection
        approaches experimentally interesting values
        $R>$1 events/kg/day.
        Applying an extensive standard scan procedure
        we have found such domains with a detection rate of about
        10 events/kg/day in $^{73}$Ge.
        We demonstrate that incorporation of the \bsg constraint  
	leads to only a moderate effect on  these domains.

        The special sampling described in
\cite{GonBer}
        has also been applied. 
	We did not reproduce large event rate domains located around 
	a value of 1~TeV for the neutralino mass, reported in the cited paper.
        No points with $R >$~1 event/kg/day have been "detected" in 
	this region of neutralino masses neither in
        the standard nor in the special scans.

\smallskip

        The paper is organized as follows.
        In section 2 we specify the MSSM and give a list
        of the formulas relevant to our analysis.
        In section 3 we summarize the experimental inputs for our
        analysis and in section 4 we summarize the formulas for 
	event rate calculations, then in section 5 discuss our 
	numerical procedure and results.
        Section 6 contains the conclusion.

\section{
	Minimal Supersymmetric Standard Model}

   	The MSSM is completely specified by the standard
   	SU(3)$\times$ SU(2)$\times$ U(1) gauge couplings as well as by
	the low-energy superpotential and "soft" SUSY breaking terms
\cite{rev}.
   	The most general gauge invariant form of the R-parity conserving 
	superpotential is
\begin{equation}
\label{superpot}
{\cal W} = h_{E}L^{j}E^{c}H_{1}^{i}{\epsilon_{ij}}
    + h_{D}Q^{j}D^{c}H_{1}^{i}{\epsilon_{ij}}
    + h_{U}Q^{j}U^{c}H_{2}^{i}{\epsilon_{ij}}
    + \mu H_{1}^{i}H_{2}^{j}{\epsilon_{ij}}
\end{equation}
  	($\epsilon_{12}=+1$). 
	The following notations are used for the quark $Q~(3,2,1/6)$,  
	$D^{c}~(\overline 3,1,1/3)$, $U^{c}~(\overline 3,1,-2/3)$, 
	lepton $L~(1,2,-1/2)$, $E^{c}~(1,1,1)$ and Higgs
	$H_{1}~(1,2,-1/2)$, $H_{2}~(1,2,1/2)$ chiral superfields with 
	the SU(3)$_c\times$SU(2)$_L\times$ U(1)$_Y$ assignment given 
	in brackets.
   	Yukawa coupling constants $h_{E, D, U}$ are matrices in the 
	generation space, non-diagonal in the general case.
   	For simplicity we suppressed generation indices.

   	In general, the ``soft'' SUSY breaking terms are given by
\cite{soft}:
\begin{eqnarray}
\label{soft}
{\cal L}_{{SB}} &=&
- {\frac {1} {2}} {\sum_{A}} M_{A} {\bar{\lambda}_{A}}{\lambda_{A}}
      -m_{H_{1}}^{2} |H_{1}|^{2}
      -m_{H_{2}}^{2} |H_{2}|^2
            -{m^{2} _{\tilde{Q}}}  |\tilde{Q}|^2 \nn  \\
      &-& {m^{2}_{\tilde{D}}} |\tilde{D^{c}}|^{2}
      - {m^{2}_{\tilde{U}}} |{\tilde{U}}^{c}|^{2} 
      - {m^{2}_{\tilde{L}}} |\tilde{L}|^{2} 
      -{m^{2}_{\tilde{E}}} |\tilde{E^{c}}|^{2} \nn \\
      &-& (h_{E} A_{E} \tilde{L}^{j} \tilde{E}^{c} H_{1}^{i}
		\epsilon_{ij}
      +h_{D} A_{D} {\tilde{Q}}^{j} {\tilde{D}}^{c} H_{1}^{i}
		\epsilon_{ij} \nn \\
      & +& h_{U} A_{U} {\tilde{Q}}^{j} {\tilde{U}}^{c} H_{2}^{i}
		\epsilon_{ij}
       +\mbox{h.c})-
      (B \mu H_{1}^{i} H_{2}^{j} \epsilon_{ij}  + \mbox{h.c})
\end{eqnarray}
   As usual, $M_{3,2,1}$ are the masses of the $SU(3)\times
   SU(2)\times U(1)$ gauginos $\tilde g, \tilde W, \tilde B$
   and $m_i$ are the masses of scalar fields.
   $A_L,\ A_D, \ A_U$ and $B$ are trilinear and bilinear couplings.

   Observable quantities can be calculated in terms of the gauge and
   the Yukawa coupling constants as well as the soft SUSY breaking
   parameters and the Higgs mass parameter $\mu$ introduced in
Eqs.~(\ref{superpot}),(\ref{soft}).
   Under the renormalization they depend on the energy scale $Q$
   according to the renormalization group equations (RGE).

   	It is a common practice to implement the grand unification (GUT)
   	conditions at the GUT scale $M_X$. 
	It allows one to reduce the  number of free parameters of the
	MSSM. 
	As explained in the
        introduction, we adopt a scenario with a non-universal Higgs mass
        with the following set of GUT conditions:
\begin{eqnarray}
\label{boundary}
 m_{\tilde L}(M_X)&=&m_{\tilde{E}}(M_X) =
   m_{\tilde Q}(M_X) = m_{\tilde U}(M_X)
                 = m_{\tilde D}(M_X) = m_{0}, \nn  \\
                 \label{bonda2}
        m_{H_{1}}^{}(M_X)&=&m_{H_{2}}^{}(M_X), \\
        A_U(M_X) &=& A_D(M_X) = A_L(M_X) = A_{0}, \nn \\
        M_{i}(M_X)&=&m_{1/2},\\
\label{gauge1}
\alpha_{i}(M_X) &=& \alpha_{GUT}, 
\mbox{ where }
\alpha_1 = \frac 53 \frac{g^{\prime2}}{4\pi},  \
\alpha_2 = \frac{g^2}{4\pi},  \
\alpha_3 = \frac{g_s^2}{4\pi},
\end{eqnarray}
        $g'~,g$ and $g_s$ are the U(1), SU(2) and SU(3) gauge coupling
	constants.
        As seen from
Eqs.(\ref{boundary}) and (\ref{bonda2}) the Higss mass parameters
        $m_{H_{1,2}}$ are not equal to the common sfermion mass $m_0$
        at the GUT scale $M_X$.
        Accepting the GUT conditions above, we end up with the
	following free MSSM parameters:
        the common gauge coupling $\alpha_{GUT}$;
        the matrices of the Yukawa couplings $h_i^{ab}$,
        where $i = E, U, D$;
        soft supersymmetry breaking parameters
        $m_0, \ m_{1/2}, \ A_0, \ B$,
        the Higgs field mixing parameter  $\mu $
        and an  additional parameter of
        the Higgs sector $m_A$ being the mass of the CP-odd neutral
        Higgs boson.
        Since the masses of the third generation are much larger
        than masses of the first two ones,
        we consider only the Yukawa coupling of the third generation
        and drop the indices $a,b$.

        Additional constraints follow from the minimization conditions
        of the scalar Higgs potential.
        Using these conditions the bilinear coupling $B$ can be
        replaced in the given list of free parameters
        by the ratio $\tb=v_2/v_1$ of the vacuum expectation values of
        the two Higgs doublets.

        We calculate the Fermi-scale parameters in
Eqs.(\ref{superpot}) and (\ref{soft})        
	in terms of the above listed free parameters on the basis of 
	2-loop RGEs following the iteration algorithm described in
\cite{WBK}.

\smallskip

        The Higgs potential $V$ including the one-loop corrections
        $\Delta V$ can be written as:
\begin{eqnarray}
  V(H_1^0,H_2^0) &=& m^2_1|H_1^0|^2+m^2_2|H_2^0|^2-
                     m^2_3(H_1^0H_2^0+ \mbox{h.c.})  \nn \\
	         &+& \frac{g^2+g^{'2}}{8}(|H_1^0|^2-|H_2^0|^2)^2 
			+ \Delta V,  \\ 
{\rm with}\quad\Delta V&=&\frac{1}{64\pi^2}
\sum_i(-1)^{2J_i}(2J_i+1)C_im_i^4
\left[\ln\frac{m_i^2}{Q^2}-\frac{3}{2}\right], \nonumber 
\label{Oneloop}
\end{eqnarray}
   	where the sum is taken over all possible particles with the
	spin $J_i$ and with the color degrees of freedom $C_i$.
   	The mass parameters of the potential are introduced in the 
	usual way as
\begin{eqnarray}
   m_{1,2}^2 = m^2_{H_{1,2}} + \mu^2, \ \ \
        m^2_3  =  B\mu,
\label{higgs_bound}
\end{eqnarray}
   	They are  running parameters with the scale Q-dependence
   	$m_i(Q)$ determined by the RGE.
   	The 1-loop potential (\ref{Oneloop}) itself is Q-independent up to,
	field-independent term depending on Q, 
   	irrelevant for the symmetry breaking.

	At the minimum of this potential the neutral components of
   	the Higgs field acquire non-zero vacuum expectation values
   	$\langle H^0_{1,2} \rangle = v_{1,2}$
   	triggering the electroweak symmetry breaking with
   	$g^2(v_1^2 + v_2^2) = 2 M_W^2$.

   	The minimization conditions read
\beq
\label{min1}
2m_1^2&=&2m_3^2\tan \beta - M_Z^2\cos 2\beta - 2\Sigma_1 \label{2m1} \\
\label{min2}
2m_2^2&=&2m_3^2\cot \beta + M_Z^2\cos 2\beta - 2\Sigma_2, \label{2m2}
\eeq
   	where 
	$\Sigma_k\equiv \frac{\partial \Delta V}{\partial \psi_k}$, 
	with
	$\psi_{1,2} = \mbox{\bf Re}H^0_{1,2}$, \ 
   	are the one-loop corrections
\cite{loopewbr}:
\begin{eqnarray}
\Sigma_k    =  -\frac{1}{32\pi^2}
\sum_i (-1)^{2J_i}(2J_i+1)\frac{1}{\psi_k}\frac{\partial m_i^2}
   {\partial \psi_k}
   m_i^2\left(\log\frac{m_i^2}{Q^2}-1\right) 
\label{sigma1}
\end{eqnarray}
        As  remnant of two Higgs doublets $H_{1,2}$ after the electroweak
        symmetry breaking there occur five physical Higgs particles:
        CP-odd neutral Higgs boson $A$, CP-even neutral Higgs bosons
        $H,h$ and a pair of charged Higgses $H^{\pm}$.
        Their masses $m_A^{}$, $m_{h,H}^{}$, $m_{H^{\pm}}^{}$
        can be calculated including all 1-loop corrections
        as second derivatives of the Higgs potential in
Eq.~(\ref{Oneloop})
        with respect to the corresponding fields evaluated at the minimum
\cite{erz,berz}.

   The neutralino mass matrix  written in the basis
   ($\tilde{B}, \tilde{W}^{3}, \tilde{H}_{1}^{0},  \tilde{H}_{2}^{0}$)
   has the form:
\begin{equation}
  {\cal M}_{\chi} = \left(
  \begin{array}{cccc}
    M_1 & 0   &-M_Zc_\beta s_w & M_Zs_\beta s_w  \\
    0   & M_2 & M_Zc_\beta c_w &-M_Zs_\beta c_w  \\
   -M_Zc_\beta s_w & M_Zc_\beta c_w & 0   & -\mu \\
    M_Zs_\beta s_w &-M_Zs_\beta c_w &-\mu & 0
  \end{array} \right), \label{neutmix} \\
\end{equation}
	where $s_w=\sin{\theta_W}$, $c_w=\cos{\theta_W}$
	and $s_\beta =\sin{\beta}$, $c_\beta=\cos{\beta}$.	   
   	Diagonalizing the mass matrix above by virtue of the orthogonal
   	matrix ${\cal N}$ one can obtain the
   	four physical neutralinos $\chi_i^{}$ with the field content
\begin{equation}
\label{admix}
\chi_i^{} = {\cal N}_{i1} \tilde{B} +  {\cal N}_{i2}  \tilde{W}^{3} +
{\cal N}_{i3} \tilde{H}_{1}^{0} + {\cal N}_{i4} \tilde{H}_{2}^{0}.
\end{equation}
   	and with masses $m_{{\chi}_i}$ being  eigenvalues of
   	the mass matrix (\ref{neutmix}).
	The lightest neutralino $\chi_1^{}$ we denote  $\chi$. 
	In our analysis $\chi$ is the lightest SUSY particle (LSP).

   The chargino  mass term is
\begin{equation}
 \left(\tilde W^-, \tilde H_1^- \right) {\cal M}_{\tilde\chi^{\pm}}
\left(  \begin{array}{c}
     \tilde W^+\\
     \tilde H_2^+ \end{array} \right)\ \  +  \ \ \mbox{h.c.}
\label{charmasster}
\end{equation}
        with the mass matrix
\begin{equation}
  {\cal M}_{\tilde\chi^{\pm}} = \left(
  \begin{array}{cc}
     M_2                  & \sqrt{2}M_W\sin\beta \\
     \sqrt{2}M_W\cos\beta & \mu
  \end{array} \right) \label{charmix}
\end{equation}
        which can be diagonalized by the transformation
\begin{eqnarray}
\label{charginos}
\tilde\chi^{-} = U_{i1}\tilde W^- +  U_{i2}\tilde H^-,   \ \
\tilde\chi^{+} = V_{i1}\tilde W^+ +  V_{i2}\tilde H^+
\end{eqnarray}
   	with $U^* {\cal M}_{\tilde\chi^{\pm}} V^{\dagger} =
  	\mbox{diag}(M_{\tilde\chi^{\pm}_1}, M_{\tilde\chi^{\pm}_2})$, 
	where the chargino masses are
\ba{charginos}
&& M^2_{\tilde\chi^{\pm}_{1,2}} = \frac{1}{2}\left[M^2_2+\mu^2+2M^2_W 
 \mp \right.  \nn \\
&& \left.
  \sqrt{(M^2_2-\mu^2)^2+4M^4_W\cos^22\beta +4M^2_W(M^2_2+\mu^2+2M_2\mu
  \sin 2\beta )}\right] \nn
\ea

  	The mass matrices for the 3-generation sfermions
   	$\tilde t$, $\tilde b$ and $\tilde\tau$ in the 
	$\tilde f_L$-$\tilde f_R$ basis are:
{\small
\ba{stopmat}
{\cal M}^2_{\tilde t}&=&\left(\begin{array}{cc}
m_{\tilde Q}^2+m_t^2+\frac{1}{6}(4M_W^2-M_Z^2)\cos 2\beta &
m_t(A_t -\mu\cot \beta ) \\ 
m_t(A_t -\mu\cot \beta ) &
m_{\tilde U}^2+m_t^2-\frac{2}{3}(M_W^2-M_Z^2)\cos 2\beta
\end{array}  \right)      
 \nonumber \\
{\cal M}^2_{\tilde b}&=&\left(\begin{array}{cc}
m_{\tilde Q}^2+m_b^2-\frac{1}{6}(2M_W^2+M_Z^2)\cos 2\beta &
m_b(A_b -\mu\tan \beta ) \\
m_b(A_b -\mu\tan \beta ) &
m_{\tilde D}^2+m_b^2+\frac{1}{3}(M_W^2-M_Z^2)\cos 2\beta
\end{array}  \right)           
    \nonumber \\
{\cal M}^2_{\tilde\tau}&=&\left(\begin{array}{cc}
m_{\tilde L}^2+m_{\tau}^2-\frac{1}{2}(2M_W^2-M_Z^2)\cos 2\beta &
m_{\tau}(A_{\tau} -\mu\tan \beta ) \\
m_{\tau}(A_{\tau} -\mu\tan \beta ) &
m_{\tilde E}^2+m_{\tau}^2+(M_W^2-M_Z^2)\cos 2\beta
\end{array}  \right)               \nonumber
\ea
}

        For simplicity we ignored in the sfermion mass matrices a
        non-diagonality in the generation space which is important
        only for the \bsg-decay.

\section{
	Constrained MSSM parameter space}

        In this section we shortly summarize the theoretical and 
	experimental constraints used in our analysis.

        Solution of the gauge coupling constants unification (see
Eq.(\ref{gauge1}))
        using 2-loop RGEs allows us to define the unification scale $M_X$.
        The following standard definitions are used:
        $\alpha_1 = 5\alpha/(3\cos^2\theta_W)$,
        $\alpha_2 = \alpha/\sin^2\theta_W$.
        The world averaged values of the gauge coup\-lings  at the
        Z$^0$ energy were obtained from a fit to the LEP data~\cite{LEP},
        $M_W$ \cite{PDB} and \mt\cite{CDF,D0}:
        $\alpha^{-1}(M_Z) = 128.0\pm0.1 $,
        $\sin^2\theta_{\overline{MS}}= 0.2319\pm0.0004$,
        $\alpha_3 =  0.125\pm0.005$.
        The value of $\alpha^{-1}(M_Z)$ was updated from 
\cite{dfs} by using new data on the hadronic vacuum polarization
\cite{EJ95}.
\smallskip
 
	SUSY particles have not been found so far and from the searches
	at LEP one knows that the lower limit on the charged sleptons
	is half the $Z^0$ ~mass (45 GeV)
	to be above 60 GeV
	For the charginos the preliminary lower limit of 65 GeV from the
	LEP 140~GeV  run was used
	The lower limit on the lightest neutralino is 18.4 GeV
	while the sneutrinos have to be above 41 GeV
\smallskip

        Radiative corrections trigger spontaneous symmetry breaking 
	in the electroweak sector.
        In this case the Higgs potential has its minimum for non-zero
        vacuum expectation values of the fields.
        Solving for $\mz$ from
Eqs. (\ref{min1}) and (\ref{min2}) yields:
\begin{equation}
\label{defmz}
\frac{\mz^2}{2}=\frac{m_1^2+\Sigma _1-(m_2^2+\Sigma _2) \tan^2\beta}
{\tan^2\beta-1},
\end{equation}
        where the $\Sigma_1$ and $\Sigma_2$ are defined in
Eq.~(\ref{sigma1}). 
        This is an important constraint which relates the true vacuum
        to the physical Z-boson mass $M_Z = 91.187\pm 0.007$GeV.

\bigskip

        Another stringent constraint is imposed by the branching ratio
        \Bbsg measured by the CLEO collaboration
\cite{cleo94}
        to be: $BR(b\to s \gamma)= (2.32\pm0.67)\times10^{-4}$.

        In the MSSM this flavour changing neutral current (FCNC) receives in
        addition to the SM\ \ $W-t$ loop contributions from $H^\pm-t$,
        $\tilde{\chi}^\pm - \tilde{t}$ and $\tilde{g}-\tilde{q}$ loops.
        The ${\chi} -\tilde{t}$ loops, which are expected to be much
        smaller, have been neglected
\cite{BerBorMasRi,borz}.
        The $\tilde{g}-\tilde{q}$ loops are proportional to $\tb$.
        It was found
\cite{WBK}
	that this contribution should  be small,
        even in the case of large \tb and therefore it was neglected.
        The chargino contribution, which becomes large for large \tb and
        small chargino masses, depends sensitively on the splitting
        of the two stop masses.

	Within the MSSM the following ratio has been calculated
\cite{BerBorMasRi}:
\ba{BSG}
&&\frac{BR(b\to s\gamma)}{BR(b\to c e \bar{\nu})}= \\
&&\frac{|V_{ts}^*V_{tb}|^2}{|V_{cb}|^2}K_{NLO}^{QCD}
\frac{6\alpha}{\pi}
\frac{\left[\eta^{16/23}A_\gamma+\frac{8}{3}(\eta^{14/23}
-\eta^{16/23})A_g+C\right]^2}{I(m_c/m_b)
[1-(2/3\pi)\alpha_s(m_b)f(m_c/m_b)]}, \nn
\ea 
	where
$$
C \approx 0.175, \ \
I = 0.4847, \ \ 
\eta=\alpha_s(M_W)/\alpha_s(m_b), \ \
 f(m_c/m_b)=2.41. \nn
$$
	Here $ f(m_c/m_b)$ represents corrections from leading order QCD
	to the known semileptonic $b\to c e \bar{\nu}$ decay rate, while the
	ratio of masses of $c$- and $b$-quarks is taken to be 
	$m_c/m_b=0.316$. 
	The ratio of CKM matrix elements
	$\frac{|V_{ts}^*V_{tb}|^2}{|V_{cb}|^2}=0.95$ was taken from 
\cite{burasb}, the next leading order QCD-corrections from 
\cite{alibsg}. 
        $A_{\gamma,g}$ are the coefficients of the
	effective operators for $bs$-$\gamma$ and for $bs$-$g$ interactions
	respectively.

\bigskip

  	 Assuming that the neutralinos  form a dominant part of
    	 the DM in the universe one obtains a cosmological constraint
   	 on the neutralino relic density.

   	 The present lifetime of the universe is at least $10^{10}$ years,
   	 which implies an upper limit on the expansion rate and
   	 correspondingly on the total relic abundance.
   	 Assuming $h_0>0.4$ one finds that  the contribution of
  	 each relic particle species $\chi$  has to obey
\bq
\label{Age_Uni}
 \Omega_\chi h^2_0<1,
\eq
   	where the relic density parameter  $\Omega_\chi = \rho_\chi/\rho_c$
   	is the ratio of the relic neutralino mass  density $\rho_\chi$
   	to  the critical one
   	$\rho_c = 1.88\cdot 10^{-29}$h$^2_0$g$\cdot$cm$^{-3}$.

   	We calculate $\Omega_{\chi} h^2_0$  following the standard
   	procedure on the basis of the approximate formula
\cite{Hagelin,drno}:
\begin{eqnarray}
\label{omega}
\Omega_{\chi} h^2_0 &=& \frac{2.13}{10^{11}}
\left(\frac{T_{\chi}}{T_{\gamma}}\right)^3
\left(\frac{T_{\gamma}}{2.7K^o}\right)^3
N_F^{1/2}
\left(\frac{{\mbox{GeV}}^{-2}}{a x_F + b x_F^2/2}\right).
\end{eqnarray}
   	Here $T_{\gamma}$ is the present day photon temperature,
   	$T_{\chi}/T_{\gamma}$ is the reheating factor,
   	$x_F = T_F/m_{\chi} \approx 1/20$, $T_F$ is
   	the neutralino freeze-out temperature, and $N_F$ is the total
   	number of degrees of freedom at $T_F$. 
	The coefficients $a, b$ are determined from the
   	non-relativistic expansion
\bq
\label{expan}
<\sigma_{annyh} v> \approx a + b x
\eq
   	of the thermally averaged cross section of neutralino annihilation.
   	We adopt an approximate treatment not taking into account
   	complications, which occur when the expansion 
(\ref{expan}) fails
\cite{omega}. 
	We take into account all possible channels of
   	the $\chi-\chi$ annihilation. 
	The most complete list of the relevant formulas for the 
	coefficients  $a, b$ and numerical values
   	for the other parameters in eqs. 
(\ref{omega}) and (\ref{expan})
   	can be found in \cite{drno}.

   	Since the neutralinos are mixtures of gauginos and
   	higgsinos, the annihilation can occur both, via
   	s-channel exchange of the $Z^0$ and Higgs bosons and
   	t-channel exchange of a scalar particle, like a selectron 
\cite{relic}.
   	This constrains the parameter space, as discussed by
   	many groups
\cite{roskane,drno,rosdm,relictst}.
   	The size of the Higgsino component
   	depends on the relative sizes of the elements
   	in the mixing matrix 
	(\ref{neutmix}), especially on \tb and the size of the
   	parameter $\mu$.

        In the analysis we ignore possible rescaling of the local
        neutralino density $\rho$ which may occur in the region of the
        MSSM parameter space where $\Omega_\chi h^2< 0.025$
\cite{BBEFMS,Gelm,Bot}.  
	This is a minimal value corresponding to DM concentrated
       	in galactic halos averaged over the universe.
        If the neutralino is accepted as a dominant part of the DM its
        density has to exceed the quoted limiting value 0.025.
        Otherwise the  presence of additional DM components should be
        taken into account, for instance, by the mentioned rescaling ansatz.  
	However, the halo density is known to be very uncertain. 
	Its actual value can be one order of magnitude smaller. 
	Therefore, one can expect that the rescaling takes place in a 
	small domain of the MSSM parameter space. 
	Another point is that the SUSY solution  of the DM problem
	with such low neutralino density becomes questionable.
        We assume neutralinos to be a dominant component of
        the DM halo of our galaxy with a density
        $\rho_{\chi}$ = 0.3 GeV$\cdot$cm$^{-3}$ in the solar vicinity
        and disregard in the analysis points with
        $\Omega_{\chi}h^2< 0.025$.

\section{Neutralino-Nucleus Elastic Scattering}

        A dark matter event is elastic scattering of a DM neutralino from
   	a target nucleus producing a nuclear recoil which can be detected 
	by a suitable detector.
        The corresponding event rate depends on the distribution of
        the DM neutralinos in the solar vicinity and
        the cross section $\sigma_{el}(\chi A)$ of  neutralino-nucleus
        elastic scattering.
        In order to calculate $\sigma_{el}(\chi A)$  one should specify
        neutralino-quark interactions.
        The relevant low-energy effective Lagrangian can be written
        in a general form as
\be{Lagr} 
  L_{eff} = \sum_{q}^{}\left( {\cal A}_{q}\cdot
      \bar\chi\gamma_\mu\gamma_5\chi\cdot
                \bar q\gamma^\mu\gamma_5 q +
    \frac{m_q}{M_{W}} \cdot{\cal C}_{q}\cdot\bar\chi\chi\cdot\bar q q\right)
      \ +\ O\left(\frac{1}{m_{\tilde q}^4}\right),
\ee
        where terms with vector and pseudoscalar quark currents are
        omitted being negligible in the case of non-relativistic
        DM neutralinos with typical velocities $v_\chi\approx 10^{-3} c$.

        In the Lagrangian \rf{Lagr} we also neglect terms which appear in
  	supersymmetric models at the order of $1/m_{\tilde q}^4$
  	and higher,  where $m_{\tilde q}$ is the mass of the scalar
        superpartner $\tilde q$ of the quark $q$.
        These terms, as recently pointed out in 
\cite{drno}, 
        are potentially important in the spin-independent 
	neutralino-nucleon scattering,
  	especially in domains of the MSSM  parameter space where
  	$m_{\tilde q}$ is close to the neutralino mass $m_{\chi}$.
  	Below we adopt the approximate treatment of these terms
        proposed in 
\cite{drno} 
	which allows "effectively" absorbing them into the
	coefficients  ${\cal C}_q$ in a wide region
  	of the SUSY model parameter space.
\ba{Aq1}
 {\cal A}_{q} = 
	&-&\frac{g_{2}^{2}}{4M_{W}^{2}} 
	   \Bigl[\frac{{\cal N}_{14}^2-{\cal N}_{13}^2}{2}T_3 \nn \\
        &-& \frac{M_{W}^2}{m^{2}_{\tilde{q}1} - (m_\chi + m_q)^2}
	   (\cos^{2}\theta_{q}\ \phi_{qL}^2 
	   + \sin^{2}\theta_{q}\ \phi_{qR}^2) \nn \\ 
        &-& \frac{M_{W}^2}{m^{2}_{\tilde{q}2} - (m_\chi + m_q)^2}
	     (\sin^{2}\theta_{q}\ \phi_{qL}^2 
	     + \cos^{2}\theta_{q}\ \phi_{qR}^2) \nn \\
        &-& \frac{m_{q}^{2}}{4}P_{q}^{2}\left(\frac{1}{m^{2}_{\tilde{q}1}
		- (m_\chi + m_q)^2}
             + \frac{1}{m^{2}_{\tilde{q}2} 
                - (m_\chi + m_q)^2}\right) \nn \\
        &-& \frac{m_{q}}{2}\  M_{W}\  P_{q}\  \sin2\theta_{q}\
            T_3 ({\cal N}_{12} - \tan\theta_W {\cal N}_{11}) \nn \\
	&\times&\left( \frac{1}{m^{2}_{\tilde{q}1}- (m_\chi + m_q)^2}
    - \frac{1}{m^{2}_{\tilde{q}2} - (m_\chi + m_q)^2}\right)\Bigr]
\ea
\ba{Cq1}
 {\cal C}_{q} = 
	&-&  \frac{g_2^2}{4} \Bigl[\frac{F_h}{m^2_{h}} h_q +
			\frac{F_H}{m^2_{H}} H_q \nn \\ 
	&+& P_q \left(\frac{\cos^{2}\theta_{q}\ \phi_{qL} -
     \sin^{2}\theta_{q}\ \phi_{qR}}{m^{2}_{\tilde{q}1} - (m_\chi + m_q)^2}
         -\frac{\cos^{2}\theta_{q}\ \phi_{qR} -
     \sin^{2}\theta_{q}\ \phi_{qL}}{m^{2}_{\tilde{q}2} - (m_\chi +
	m_q)^2}\right) \nn \\
	&+& \sin2\theta_{q}(\frac{m_q}{4 M_W} P_{q}^{2} -
                           \frac{M_W}{m_q} \phi_{qL}\ \phi_{qR}) \nn \\
	&\times&\left(\frac{1}{m^{2}_{\tilde{q}1} - (m_\chi + m_q)^2} -
\frac{1}{m^{2}_{\tilde{q}2} - (m_\chi + m_q)^2}\right)\Bigr].
\ea
     Here
\ba{Fq1}
        F_{h} &=& ({\cal N}_{12} - {\cal N}_{11}\tan\theta_W)
        ({\cal N}_{14}\cos\alpha_H + {\cal N}_{13}\sin\alpha_H), \nn \\
        F_{H} &=& ({\cal N}_{12} - {\cal N}_{11}\tan\theta_W)
        ({\cal N}_{14}\sin\alpha_H - {\cal N}_{13}\cos\alpha_H),\nn \\
h_q &=&\bigl(\frac{1}{2}+T_3\bigr)\frac{\cos\alpha_H}{\sin\beta}
          - \bigl(\frac{1}{2}-T_3\bigr)\frac{\sin\alpha_H}{\cos\beta},\nn \\
H_q &=& \bigl(\frac{1}{2}+T_3\bigr)\frac{\sin\alpha_H}{\sin\beta}
      + \bigl(\frac{1}{2}-T_3\bigr)\frac{\cos\alpha_H}{\cos\beta}, \nn \\
\phi_{qL} &=& {\cal N}_{12} T_3 + {\cal N}_{11}(Q -T_3)\tan\theta_{W},\nn \\
\phi_{qR} &=& \tan\theta_{W}\  Q\  {\cal N}_{11}, \nn \\
P_{q} &=&  \bigl(\frac{1}{2}+T_3\bigr) \frac{{\cal N}_{14}}{\sin\beta}
          + \bigl(\frac{1}{2}-T_3\bigr) \frac{{\cal N}_{13}}{\cos\beta} \nn
\ea
        Our formulas for the coefficients ${\cal A}_q$ and ${\cal C}_q$
        of the effective Lagrangian take into account squark mixing
   	$\tilde{q}_L-\tilde{q}_R$ and the contribution of
   	both CP-even Higgs bosons $h, H$.
        The formulas coincide with the relevant formulas in  
\cite{drno} 
	neglecting the terms $\sim 1/m_{\tilde q}^4$ and higher.
        These terms are taken into account "effectively" by
   	introducing an "effective"  stop quark $\tilde t$
        propagator.

        A general representation of the differential cross section
        of neutralino-nucleus scattering can be given in terms of
        three spin-dependent ${\cal  F}_{ij}(q^2)$ and
        one spin-independent ${\cal F}_{S}(q^2)$ form factors as follows
\cite{EV}
\ba{cs}
\dsdq2(v,q^2)&=&\frac{8 G_F}{v^2} \Bigl(
   		a_0^2\cdot {\cal F}_{00}^2(q^2) 
	     + a_0 a_1 \cdot {\cal F}_{10}^2(q^2) \nn \\
	     &+&
   	        a_1^2\cdot {\cal F}_{11}^2(q^2)
   	      + c_0^2\cdot A^2\ {\cal F}_{S}^2(q^2)
   	\Bigr).
\ea
   	The last term corresponding to the spin-independent scalar 
	interaction gains coherent enhancement $A^2$  
	($A$ is the atomic weight of the nucleus in the reaction).
   	The coefficients $a_{0,1}, c_0$ do not depend on nuclear structure
   	and relate to the parameters ${\cal A}_q$, ${\cal C}_q$ 
	of the effective Lagrangian 
\rf{Lagr} and to the parameters $\Delta q$, $f_s$, $\hat{f}$
   	characterizing the nucleon structure. One has the relationships
\ba{rel1}
a_0&=&( {\cal A}_u + {\cal A}_d)
        ( \Delta u + \Delta d ) + 2 \Delta s {\cal A}_s, \nn \\
a_1&=&({\cal A}_u - {\cal A}_d)
        (\Delta u - \Delta d ),   \\ 
c_0&=&\hat f \frac{m_u {\cal C}_{u}
        + m_d {\cal C}_{d}}{m_u + m_d}
+ f_{s} {\cal C}_{s}  + \frac{2}{27}(1- f_{s} - \hat f)({\cal C}_{c}
        + {\cal C}_{b} + {\cal C}_{t}). \nn
\ea
   	Here  $\Delta q^{p(n)}$ are the fractions of the 
	proton(neutron) spin carried by the quark $q$.
        The standard definition is
\be{Spin} 
   <p(n)|\bar q\gamma^\mu\gamma_5 q|p(n)> = 2 S_{p(n)}^{\mu} \Delta q^{p(n)},
\ee
        where $S_{p(n)}^{\mu}=(0,\vec{S}_{p(n)})$ is the 4-spin of 
	the nucleon.
        The parameters $\Delta q^{p(n)}$ can be extracted from data 
	on polarized nucleon structure functions
\cite{EMC,SMC2} and hyperon semileptonic decay data
\cite{Dqextract}.

        We use in the analysis $\Delta q$ values extracted both from the EMC
\cite{EMC} and from SMC
\cite{SMC2} data.

        The other nuclear structure parameters $f_s$ and $\hat f$
        in formula 
\rf{rel1} 
	are defined as follows:
\ba{Scal}
   <p(n)|(m_{u} + m_{d})(\bar{u}u + \bar{d}d)|p(n)> &=& 2\hat f M_{p(n)}
           \bar \Psi \Psi, \\
\nn
   <p(n)|m_{s}\bar{s}s|p(n)> &=& f_{s} M_{p(n)}\bar \Psi \Psi.
\ea
        The values extracted from the data under certain theoretical
        assumptions are
\cite{ChengGasser}:
        \ba{f}
   \hat{f} = 0.05\ \ \ \ \ \ \mbox{and}\ \ \ \ \ \ f_{s} = 0.14.
        \ea
   	The strange quark contribution $f_{s}$ is known to be uncertain
        to about a factor of 2. 
	Therefore we take its value in the analysis
        within the interval $ 0.07 < f_{s} < 0.3$
\cite{Hatsuda,ChengGasser}.

        The nuclear structure comes into play via the form factors
        ${\cal F}_{ij}(q^2), {\cal F}_{S}(q^2)$ in Eq. \rf{cs}.
        The spin-independent
        form factor ${\cal F}_{S}(q^2)$ can be represented as the normalized
        Fourier transform of a spherical nuclear ground state density
   	distribution $\rho({\bf r})$.
   	In the analysis we use the standard Woods-Saxon inspired
   	distribution
\cite{E}.
   	It leads to the form factor
\be{fourier}
{\cal F}_{S}(q^2) = \int d^3{\bf r} \rho({\bf r}) e^{i {\bf r q}}
= 3\frac{j_1(q R_0)}{q R_0} e^{-\frac{1}{2} (qs)^2},
\ee
   	where $R_0 = (R^2 - 5 s^2)^{1/2}$ and $s \approx 1$ fm are the
   	radius and the thickness of a spherical nuclear surface 
	respectively, $j_1$ is the spherical Bessel function of index 1.

   	Spin-dependent form factors ${\cal F}_{ij}(q^2)$ are much more
   	nuclear model dependent quantities.
   	The last few years have seen a noticeable progress
   	in detailed nuclear model calculations of these form factors.
   	For many nuclei of interest in DM search  they have been calculated
   	within the conventional shell model
\cite{Ressell} and within an approach based on the theory of 
	finite Fermi systems
\cite{Nikolaev}.
   	We use the simple parameterization of the  $q^2$ dependence of
   	${\cal F}_{ij}(q^2)$ in the form of a Gaussian with the 
        {\em r.m.s.} spin radius
   	of the nucleus calculated in the harmonic well potential
\cite{EF}.
   	For our purposes this semi-empirical scheme is sufficient.

        An experimentally observable quantity is the differential event rate
        per unit mass of the target material
\be{drate1}
\frac{dR}{dE_r} = \preRate
    \int^{v_{max}}_{v_{min}} dv f(v) v \dsdq2 (v, E_r)
\ee
   	Here $f(v)$ is the velocity distribution of neutralinos
   	in the earth's frame which is usually assumed to be a 
	Maxwellian distribution in the galactic frame.
$v_{max} = v_{esc} \approx$ 600 km/s and
$\rho_{\chi}$ = 0.3 GeV$\cdot$cm$^{-3}$  are the escape velocity
	and the mass density of the relic neutralinos in
   	the solar vicinity;
$v_{min} = \left(M_A E_r/2 M_{red}^2\right)^{1/2}$ with
$M_A$ and $M_{red}$ being the mass of nucleus $A$ and the reduced
   	mass of the neutralino-nucleus system, respectively. 
	Note that $ q^2 = 2 M_A E_r$.

   	The differential event rate is the most appropriate quantity	
	for comparing with the observed recoil spectrum and allows one
	to take properly into account spectral characteristics of a 
	specific detector and to separate the background.
   	However, in many cases the total  event rate $R$\
   	integrated over the whole kinematical domain of the recoil energy
   	is sufficient.
   	It is widely employed in theoretical papers for estimating the
	prospects for DM detection,
   	ignoring experimental complications which may occur on the  way.
   	Notice, that the integrated event rate is less sensitive
   	to details of nuclear structure then the differential one
\rf{drate1}.
   	The $q^2$ shape of the form factors ${\cal F}_{ij}(q^2),
   	{\cal F}_{S}(q^2)$ in
Eq. \rf{cs} 
	may essentially change from one nuclear model to another.
   	Integration over $q^2$ as in the case of
   	the total event rate $R$\ reduces this model dependence.
        In the present paper we are going  to perform a general 
	analysis aimed at searching for domains with extraordinary 
	large values of the event rate $R$\ like those reported in
\cite{GonBer}.
   	This is the reason why we use in the analysis the total 
	event rate $R$.

\section{Numerical Analysis}

   	In our numerical analysis we randomly  scan the MSSM parameter
   	space within a broad domain
\ba{domainn}
 1\ GeV < m_{1/2} &<& 5\ TeV,\ \ \ \ \  \ \ \ \ \ \ \ \ \ \ \ \ \ \
                            |\mu| < 2\ TeV, \\
1 < \tan\!\beta &<& 50, \ \ \ \ \ \ \ \ \ \ \ \ \ \ \ \
                          \ \ \ \ \ \ |A_0| < 1\ TeV,\\
\label{domain} 0 < m_0 &<& 5\ TeV, \ \ \ \ 50\ GeV < m_A < 1\ TeV.
\ea

  	In the region where $\tan\!\beta \gsim 35$ the top Yukawa dominance
  	approximation is not applicable in the RGE. Therefore, we use
  	the procedure developed in
\cite{WBK} 
	which takes into account the bottom and tau Yukawa couplings as well.

   	The cut-off condition $R >$ 0.01 event/kg/day  
        is implemented in the scanning procedure.
	It reflects realistic sensitivities of the present 
	and the near-future DM  detectors.

        Note again, that we use the GUT scenario with
        the non-universal Higgs mass parameters (see 
(\ref{boundary}))
\cite{BKK}.
        The Higgs boson masses are calculated in terms of
        the CP-odd Higgs boson  mass $m_A$ and other input parameters.
        If we adopt the ultimate GUT conditions with all scalar
        mass parameters being equal at the unification scale $M_X$,
        the CP-even Higgs boson mass  becomes too big because of
        strong correlations with the sfermion spectrum.
        As a result the total event rate $R$\ decreases to  small values,
        typically less than 0.01 event/kg/day.
\smallskip

   	The main results of our scan are presented in
	Figs.1-4 in the form of scatter plots.
   	Given in
Figs.1-3 are the total event rates $R$ for $^{73}$Ge, 
   	Al$_2$O$_3$, and  NaI versus neutralino mass $m_\chi$, 
	as well as $R$ versus the ratios $R_{sd}/R$ of
   	the corresponding spin-dependent ($R_{sd}$)
   	part of $R$ to $R$ ($R=R_{sd}+R_{si}$).
Fig. 4 presents the neutralino relic density $\Omega_{\chi}h^2$
   	as a function of $m_{\chi}$.
   	All quantities are given with and without the \bsg constraint.

\smallskip

   	We find that the \bsg constraint strongly reduces the 
	MSSM parameter space.
  	The restriction leaves about 25\% of the points of the MSSM
   	parameter space which successfully have passed all 
	other constraints.

   	This constraint disfavors negative values (in our notation)
   	of the Higgs mixing parameter $\mu$. 
	It also shrinks the allowed domain for the parameter $m_{1/2}$ and
   	consequently reduces the allowed domain for the 
	LSP mass $m_{{\chi}}$.

   	As seen from
Figs.1-3 the \bsg constraint strongly suppresses those points 
	in the parameter space which correspond to the spin-dominant 
	($R_{sd}>R_{si}$) event rates in all isotopes analyzed.
	The observation strengthens 
	the conclusion about dominance of 
	the spin-independent neutralino interaction with nuclei 
	obtained in 
\cite{BKK} without \bsg constraint.
\smallskip

   	Nevertheless it is clear that the large event rates survive
   	the \bsg constraint.
        In the table~1 we present 5 examples of large event rate
	points taken from the scatter plots in Fig.~1.

\bigskip

        In paper
\cite{GonBer} 
	extraordinary large event rates of about 10 events/kg/day 
	for $^{76}$Ge and 100 events/kg/day for NaI were found
        in a specially arranged scan in the domain
        800~GeV $<m_{\chi}<$ 1200~GeV,\  0.01$<Z_g<$0.99, \  
	0$<m_A<$60~GeV
        ($Z_g = {\cal N}_{11}^2 + {\cal N}_{12}^2$ \ see 
Eq. (\ref{admix})).
        We have thoroughly scanned this region to check the cited
        striking result.
        We arrived at a negative conclusion.
        No large event rate domains around
        $m_{\chi} \sim 1$ TeV as quoted in 
\cite{GonBer} 
	have been found in our scan.
        Note, since the neutralino is the LSP, these domains correspond
        to a situation when all SUSY particles are very heavy with
        masses around 1 TeV or larger.
        Looking at the formulas 
\rf{Aq1} and \rf{Cq1} 
	we do not see any natural
        possibility for $R$\ to approach such large values in this domain. 
        The strong kinematical suppression can only be compensated
        in the case when $m_{\tilde{q}} - m_\chi \simeq m_q$.

\bigskip

\section{Conclusion}

        We have systematically studied the allowed MSSM parameter 
	space taking into account various theoretical and experimental
        constraints.
        We have found domains with experimentally interesting
        event rates for the DM neutralino detection
        ($R\simeq$ 10 events/kg/day) in the neutralino mass range
        70~GeV $<m_\chi<$ 200~GeV.
        This would be within the reach of current dark matter experiments.
        Special attention was paid to the constraint following from
        the CLEO measurement of \Bbsg.
        We have illustrated that despite the well known fact that
        this constraint essentially reduces the allowed MSSM parameter
        space it does not exclude large event rate domains. 
        We have checked the recently reported result
\cite{GonBer} on large neutralino detection event
        rates in the 1 TeV region of the neutralino mass. 
	Our analysis has not reproduced this result.

\bigskip
\newpage

{\bf  Acknowledgments}

\smallskip

   The authors wish to thank professors W. de Boer and
   D.I.Kaza\-kov, Drs R.Ehret and W.Obersulte-Beckmann
   for fruitful discussions and help with calculations.

\bigskip  

%

\newpage 

{}\vspace*{2cm}

\centerline{\bf Figure legends}

{}\vspace*{1cm}

\begin{center}

\vbox{
{\hsize 4.0in
	
\begin{itemize}
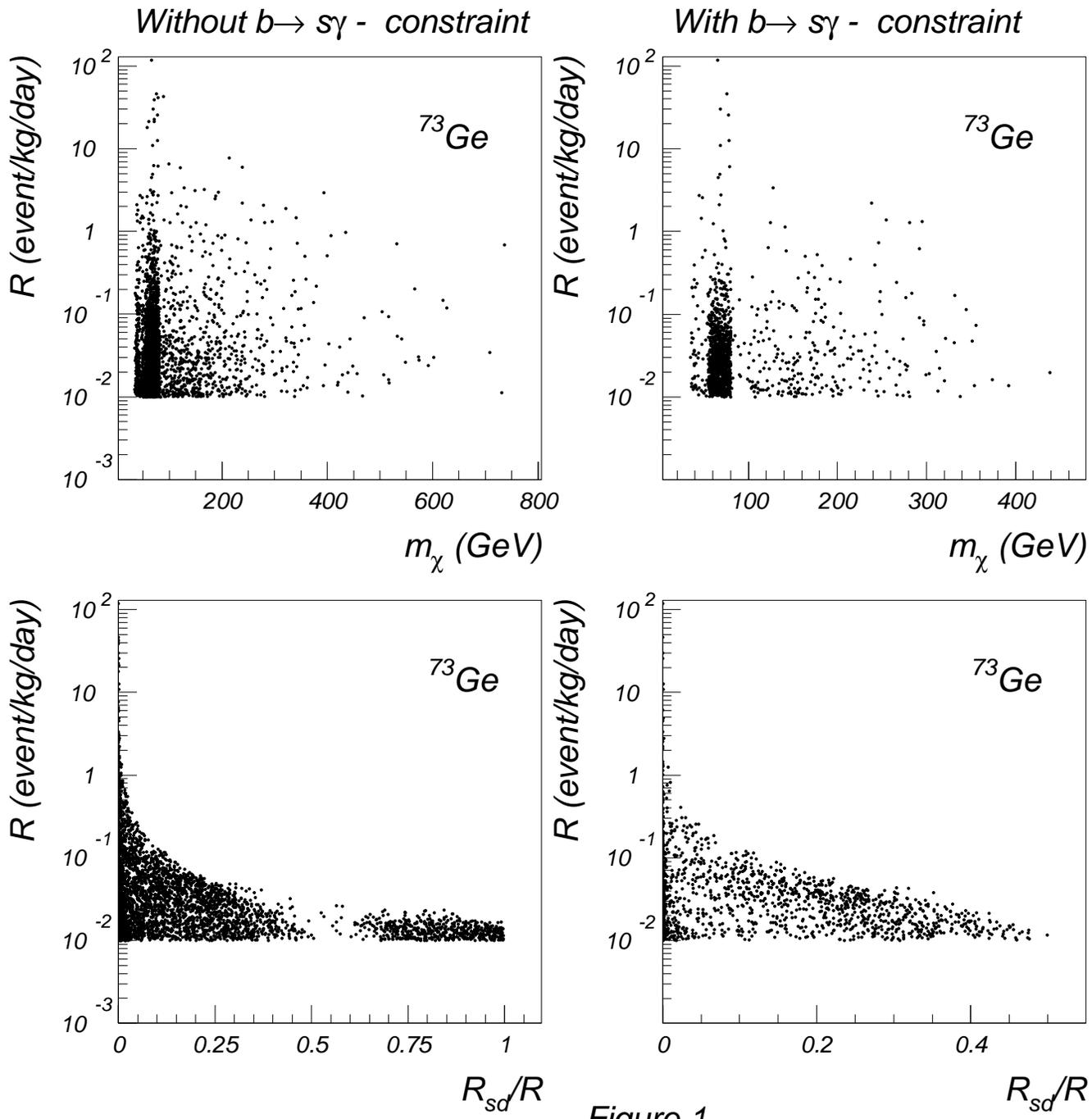
\item[Figure 1.]
        The total event rate $R$\ for  $^{73}$Ge,
        versus mass of neutralino $m_\chi$ (upper panel)
        as well as versus the ratio $R_{sd}/R$\  of
        the corresponding spin-dependent $R_{sd}$\ part of $R$\
        to the $R$\ ($R=R_{sd}+R_{si}$).
        The scatter plots are obtained without (left panel) and
        with the \bsg constraint (right panel).
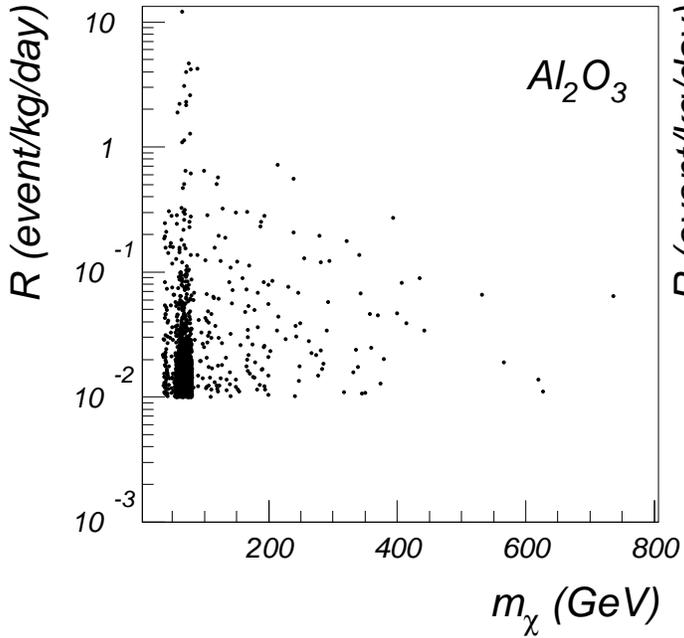
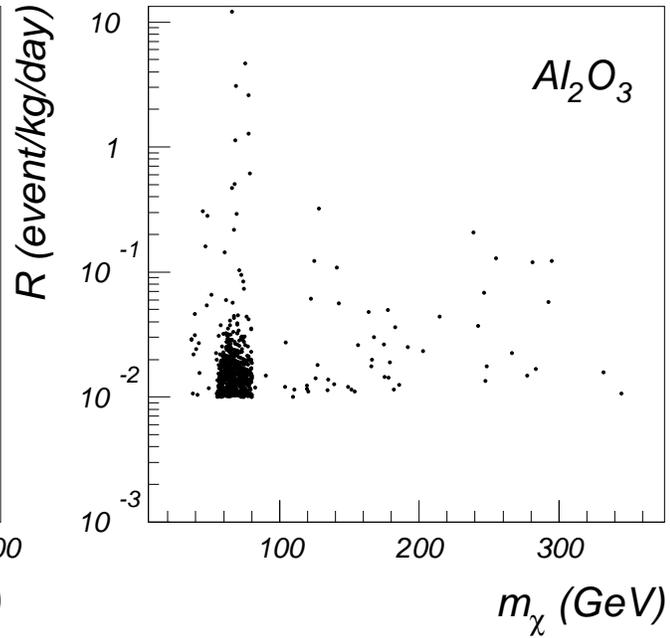
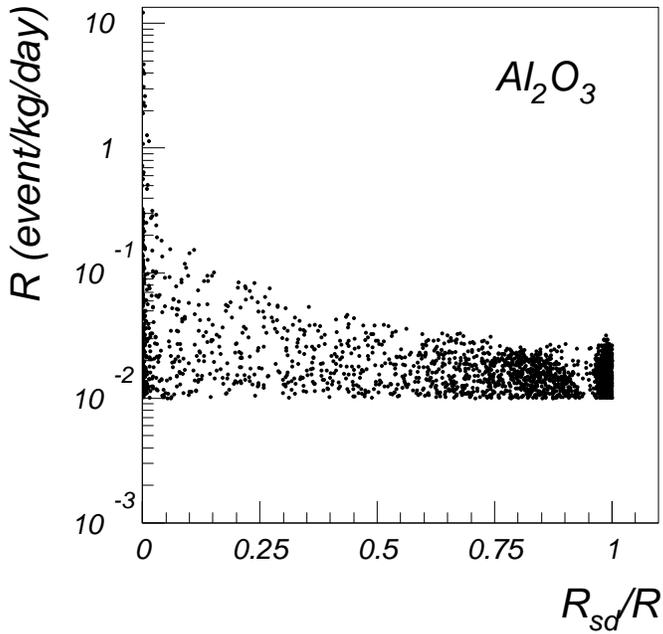
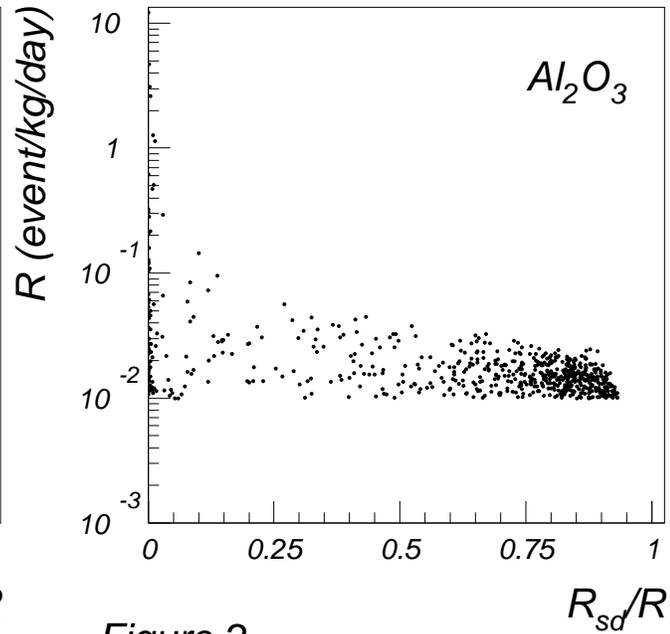
\item[Figure 2.]
	  The same as in Figure 1, but for sapphire, Al$_2$O$_3$.
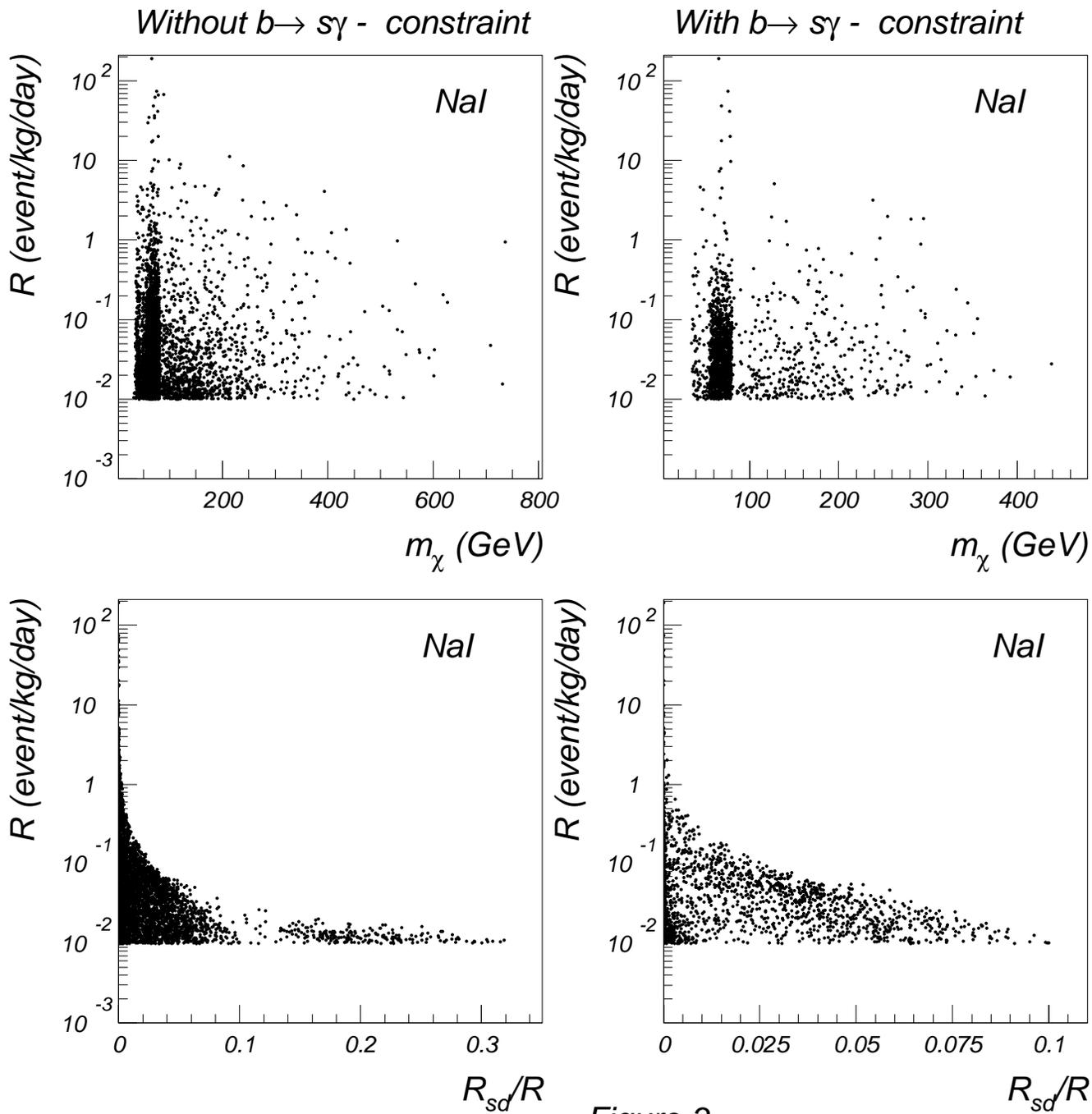
\item[Figure 3.]
    	The same as in Figure 1, but for sodium iodide, NaI.
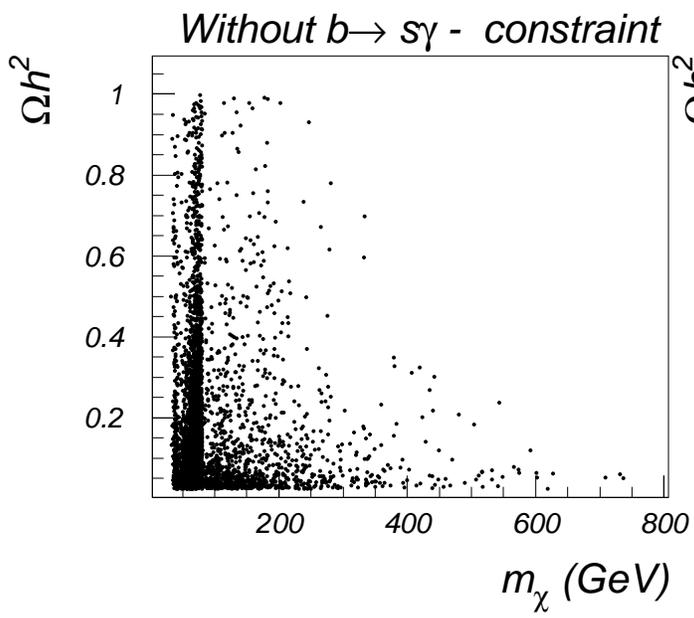
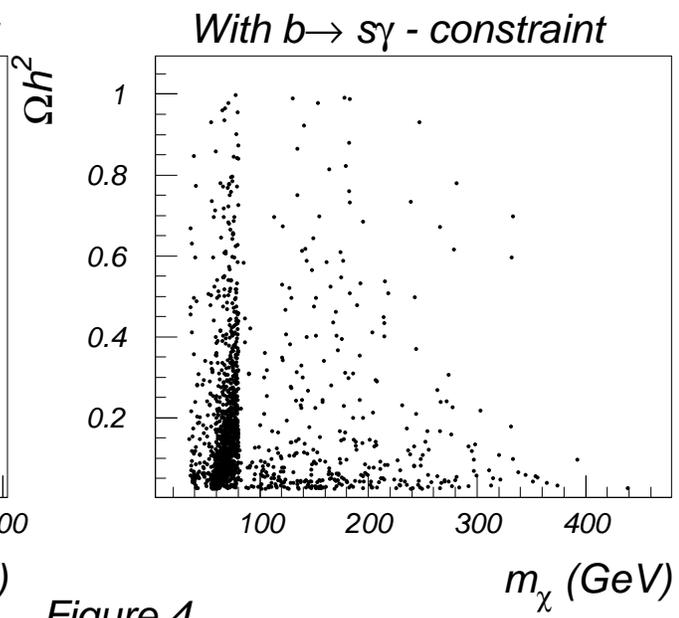
\item[Figure 4.]
        The relic density $\Omega_{\chi}h^2$
        versus mass of neutralino $m_\chi$
        without (left panel) and
        with the \bsg constraint (right panel).
\end{itemize}

}}

\end{center}

\newpage  

{}\vspace*{2cm}

\begin{table}

\centerline{\bf Table} 

{}\vspace*{1cm}

\vbox{
\begin{center}
\begin{tabular}{|l|rrrrr|} \hline
SUSY points    &  1     &   2      &   3    &   4    &  5   \\ \hline
$\tan\beta$           & 20.4   &  21.2    & 21.2   & 12.7   & 19.5 \\
$m_0$~(GeV)             & 3654   &  1421    & 3055   & 646    & 590  \\
$m_{1/2}$~(GeV)         & 621    &  229   & 405    & 372    & 320  \\
$A_0$~(GeV)             & -2.8   & -0.18  & 0.5    & -5.3   & -1.4 \\
$m_A$~(GeV)             & 941    & 588 & 673    & 575    & 685  \\
$\mu$~(GeV)             & 176    & 575 & 678    & 606    & 652  \\ \hline
$m_{\chi}$~(GeV)& 70.1   & 91.3   & 163    & 148    & 129  \\ \hline
$\Omega_\chi h^2_0$   & 0.17   & 0.074 &  0.1   & 0.054  & 0.16 \\ \hline
R(events/kg/day)      & 7.38   & 1.08  & 2.13   & 2.25   & 1.73 \\  \hline
$R_{sd}/R_{si}\cdot 10^3$& 0.5 & 0.2   & 0.04& 0.07& 0.06 \\ \hline
Gaugino fraction, $Z_g$  & 0.04   & 0.95  & 0.95   & 0.93   & 0.96  \\ \hline
\Bbsg$\cdot 10^{3}$   & 0.28  & 0.22& 0.29& 0.25& 0.19 \\ \hline
\end{tabular}
\end{center}

Table 1. Representative points with large event rate values for 
	germanium, $^{73}$Ge.
        (The gaugino fraction is defined as 
        $Z_g={\cal N}_{11}^2 + {\cal N}_{12}^2$)

}
\end{table}

\end{document}